\documentclass[12pt,a4paper]{article}
\usepackage[T1]{fontenc}
\usepackage{fancyhdr}
\usepackage{amsmath,amsthm,amsfonts,amssymb}
\allowdisplaybreaks% page breaks allowed in align equations
\usepackage{epic,eepic}
\usepackage{graphicx}
\usepackage{authblk}
\usepackage{hyperref}
\usepackage{fullpage}
\usepackage[active]{srcltx}
\usepackage{enumitem}% itemize options
\usepackage{booktabs}% for tables
\usepackage{mathtools}% this is required to write text over \rightarrow

\usepackage{framed,color}

\usepackage{float}%
\floatstyle{plaintop}%
\restylefloat{table}% these three lines put caption for tables on top

\usepackage[font={small,it}]{caption}% the options makes the font italic in floats
\usepackage[table]{xcolor}% use colours in table rows

\usepackage[toc,page]{appendix}
\graphicspath{{Figures/}}

\usepackage[style=authoryear,backend=bibtex,maxcitenames=2,maxbibnames=99]{biblatex}
\bibliography{biblio}

\usepackage{chngcntr}
\usepackage{dsfont}% indicator function \mathds{1} symbol

\usepackage{algpseudocode}
\usepackage{algorithm}
\usepackage{setspace}% for the pseudocode

\usepackage{tikz}
\usetikzlibrary{bayesnet}
\usetikzlibrary{decorations.pathreplacing}%decoration library

\makeatletter
\AtEveryBibitem{%
  \global\undef\bbx@lasthash%
  \clearfield{}}
\makeatother

\theoremstyle{plain}

\theoremstyle{definition}

\theoremstyle{remark}

\numberwithin{equation}{section}
\numberwithin{figure}{section}
\theoremstyle{plain}

\usepackage{color}

\date{}
\title{\textbf{Discussion of the article "Bayesian cluster analysis: point estimation and credible balls" by Wade and Ghahramani}}
\author[1]{Nial Friel}
\author[2]{Riccardo Rastelli}
\affil[1]{\footnotesize School of Mathematics and Statistics and Insight Centre for Data Analytics, University College Dublin, Ireland; nial.friel@ucd.ie;}
\affil[2]{\footnotesize Institute for Statistics and Mathematics, Vienna University of Economics and Business, Austria; riccardo.rastelli@wu.ac.at.}

\begin{document}
\baselineskip=20pt
\counterwithout{figure}{section}
\counterwithout{figure}{subsection}
\counterwithout{equation}{section}
\counterwithout{equation}{subsection}

\maketitle
\begin{abstract}
\noindent
We present a discussion of the paper ``Bayesian cluster analysis: point estimation and credible balls'' by \textcite{wade2018bayesian}.
We believe that this paper contributes substantially to the literature on Bayesian clustering by filling in an important methodological gap, by
providing a means to assess the uncertainty around a point estimate of the optimal clustering solution based on a given loss function. 
In our discussion we reflect on the characterisation of uncertainty around the Bayesian optimal partition, revealing other possible alternatives that may be viable.
In addition, we suggest other important extensions of the approach proposed which may lead to wider applicability. 
\\

\noindent
{\bf Keywords:}
Bayesian clustering; Greedy optimisation; Latent variable models; Markov chain Monte Carlo.
\end{abstract}

We congratulate the authors, Wade and Ghahramani (W\&G hereafter), on a wonderful article \parencite{wade2018bayesian} which is an 
excellent contribution to the area of Bayesian cluster analysis. Here the authors address the problem of appropriately summarising a partition based on a posterior.
This is a crucial issue arising in a variety of clustering contexts. 
While Markov chain Monte Carlo techniques, for example, can be used to efficiently sample the cluster membership variables from the posterior distribution of a variety of mixture models, 
it is not immediately clear then how one can reasonably summarise  such information.
Similarly to other previous papers, notably \textcite{lau2007bayesian}, the authors define the optimal partition as the one minimising the posterior expectation of a suitable loss function, and propose a 
greedy algorithm to estimate such an optimal solution. Somewhat surprisingly, there has been very little in the literature around how one might assess the uncertainty in this point estimate. W\&G address
this crucial gap by introducing a strategy to characterise the uncertainty around the optimal partition using an adaptation of the credible 
intervals approach. We consider this to be a major contribution and expect it stimulate future developments in this field. 

We have recently worked on the same problem and published our findings in \textcite{rastelli2017optimal} (hereafter referred to as R\&F).
Similarly to W\&G, we rely on a decision theoretic framework to summarise a collection of partitions, however, differently from their approach, our contribution is primarily focused on the computational 
aspects of the problem. Our method is implemented in the \texttt{R} package \texttt{GreedyEPL} available on \texttt{CRAN}.
In this discussion we compare our findings to those of W\&G mainly focusing on the following aspects: the choice of loss function used and the ensuing computational
complexity; alternatives to the credible balls approach; the wider applicability of the methods proposed.

\section{Choice of loss function and computational efficiency}
As W\&G clearly point out in their paper, commonly used loss functions such as the $0-1$ loss or the squared error loss are not ideally suited to compare partitions, due to the discrete nature of the variables 
and because of the lack of total order in the space. This leads to the important issue of finding an appropriate and reasonable loss function to compare partitions.
A popular choice in this context is Binder's loss, primarily for two main reasons: 
its close connection to the Rand index; but also since the corresponding optimal partition can be estimated via the posterior similarity matrix, which itself can be routinely estimated by
MCMC, for example. The posterior similarity matrix is an $N\times N$ matrix with element $n,n'$ (denoted $p_{n,n'}$ in W\&G) equal to the posterior probability that observations $n$ and $n'$ are 
allocated to the same cluster and where $N$ denotes the size of the dataset.
The Variation of Information (VI) loss does not possess such a representation in terms of the posterior similarity matrix and as such it turns out that this 
brings with it an increased computational overhead. 
However, W\&G neatly sidestep this problem by exploiting Jensen's inequality to obtain a lower bound for the VI loss which relies only on the posterior similarity matrix.
This input is interesting, though we note that the effect that this approximation has on the estimated optimal partition is not clear.

In R\&F, the approach we advocate does not rely on the posterior similarity matrix representation and does not involve any approximation.
In fact, our method may be used with any loss function, $\mathcal{L}\left( \textbf{a},\textbf{z} \right)$ that depends on the the two partitions, $\textbf{a}$ and $\textbf{z}$ 
through the counts $n_{ij}$, denoting the number of data points allocated to group $i$ in partition $\textbf{a}$ and to group $j$ in partition $\textbf{z}$, which can conceptually 
be considered as depending on the contingency table defined by both partitions.
Binders' loss and VI loss are included in this family, along with other known losses such as the normalised VI and the normalised information distance.
Moreover, since our approach does not require the posterior similarity matrix, its computational complexity in $N$ is decreased to a linear order (See Figure~$1$ of \textcite{rastelli2017optimal}). 
However, the computational cost of our approach also becomes increasingly costly as sample size of partitions drawn from the posterior increases. 

Additionally, R\&F empirically assess the effect of the various loss functions on simulated data and in particular we refer the reader to Figure $3$ of \textcite{rastelli2017optimal}. The main take 
home message is that the VI loss typically
achieved the best results in terms of the number of estimated groups, while the other loss functions, including Binders loss, the normalised VI loss and
the normalised information distance often exhibit unreasonable behaviour and overestimation of the number of groups. 
However, our findings also reveal that the VI loss tends to be biased towards an overestimation of the number of groups.
This seems not to be case with the results presented in W\&G. We wonder if the approximation the authors introduce may have an impact on the estimation of 
the number of groups? All things considered, we deem the research question of finding an optimal loss function and associated computational strategy 
still very open.

\section{Quantifying the uncertainty around the estimated Bayes partition}
In our experience, the marginal posteriors for the cluster membership variables generally exhibit some degree of multimodality, even after labeling issues have been 
taken into account. This is one important reason why often Markov chain Monte Carlo sampling methods generally struggle to explore the discrete search space efficiently.

\begin{figure}
\includegraphics[width = 0.49\textwidth]{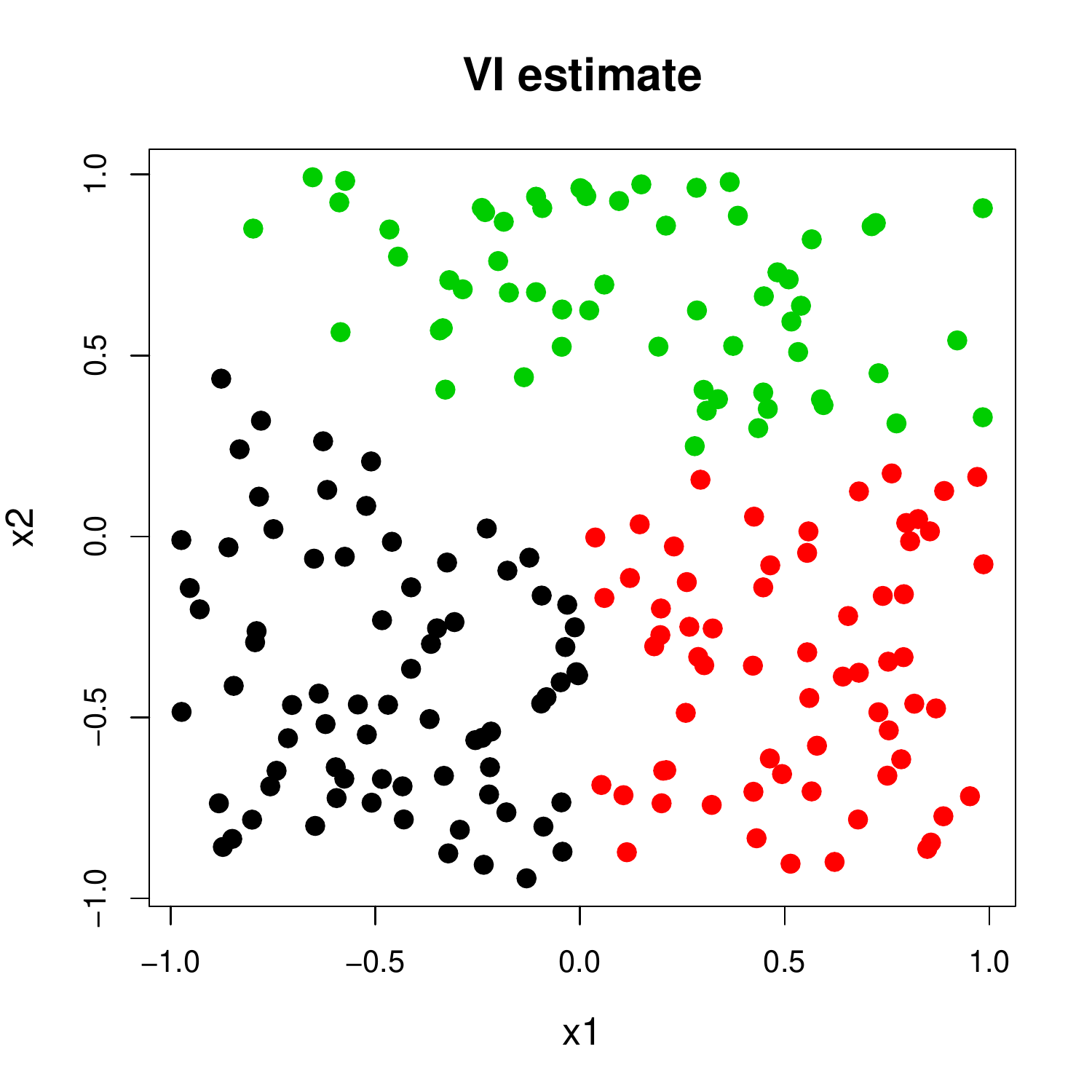}
\includegraphics[width = 0.49\textwidth]{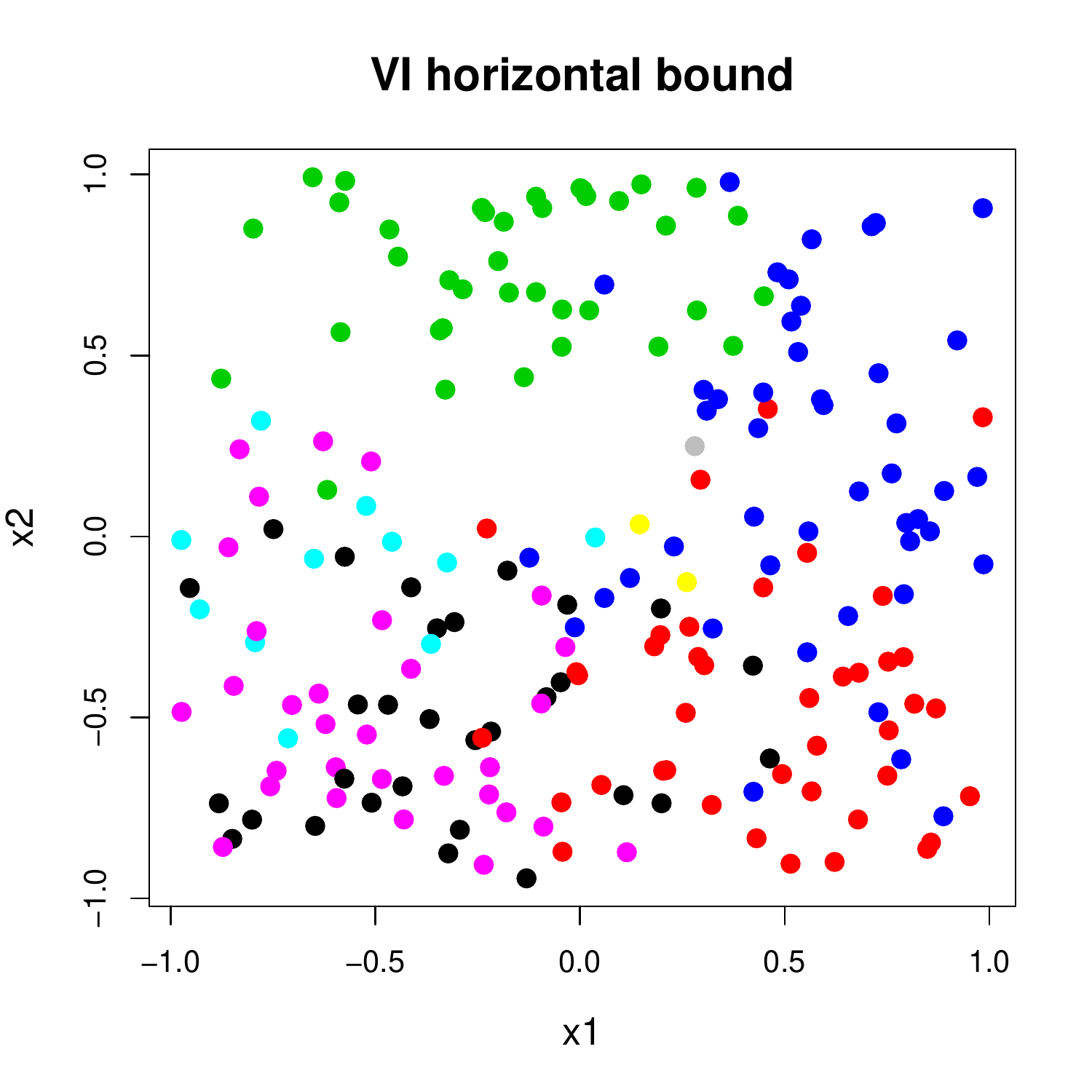}\\
\includegraphics[width = 0.49\textwidth]{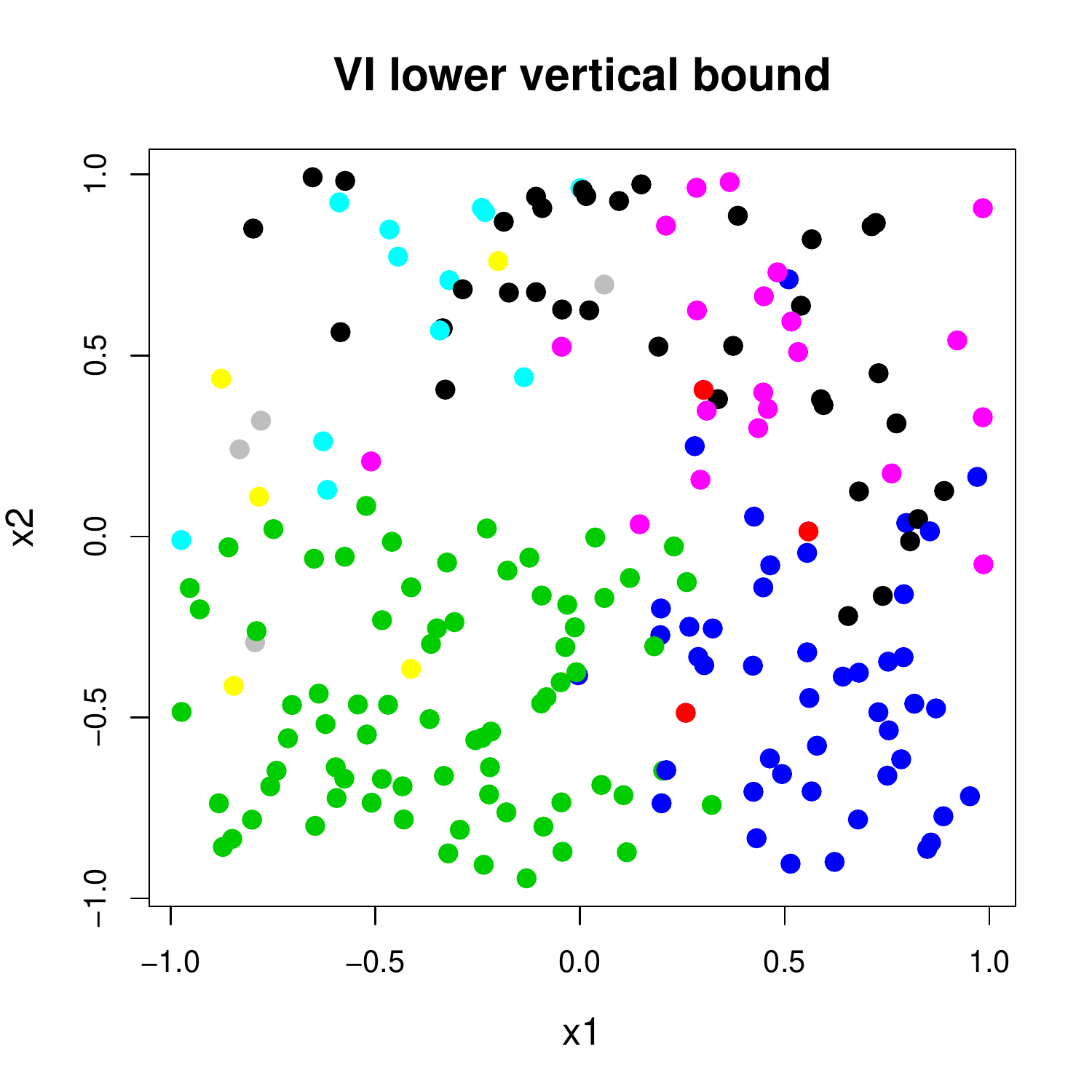}
\includegraphics[width = 0.49\textwidth]{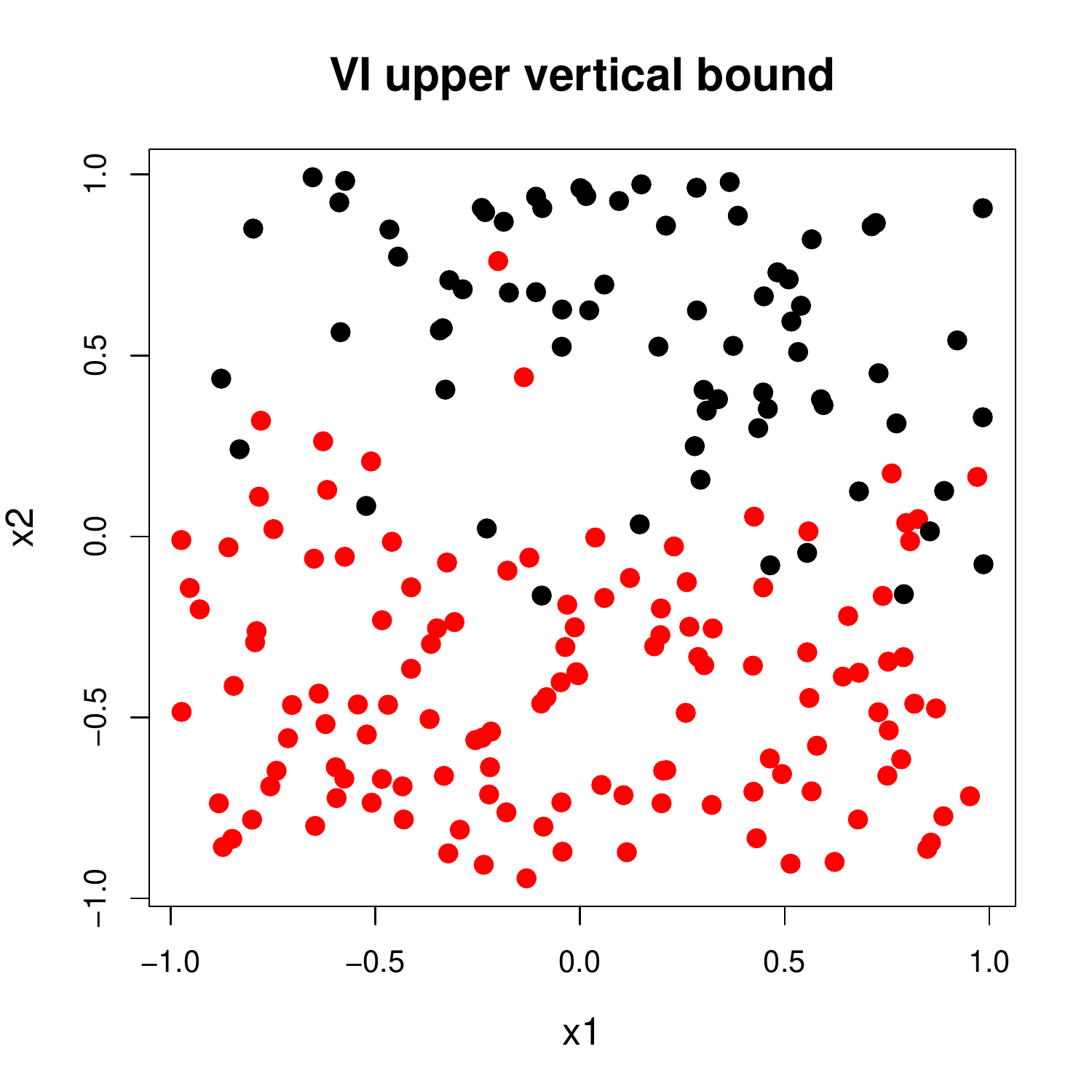}
\caption[]{VI loss optimal clustering and credible ball bounds for the simulated uniform data proposed.}
\label{fig:simulations_2}
\end{figure}

We believe that the same multimodality may also have non-negligible effects on the characterisation of the uncertainty around the estimated optimal Bayes partition.
In a nutshell: if, by definition, the credible ball has to include $95\%$ of the posterior mass, it will contain most of the relevant modes, but it may also include 
many of irrelevant partitions ``between'' them, in the sense of the loss considered. This would result in a quite heterogeneous set which may be hard to characterise, 
and where the horizontal and vertical bounds may not be so relevant to the clustering problem. We present a small experiment here to illustrate this point. 
Here we simulated a data set by sampling from a uniform distribution in the square $[-1,1]\times[-1,1]$.
We assumed the data followed a Gaussian mixture model and then obtained a posterior sample of partitions using the \texttt{R} package \texttt{bayesm}. We then applied the methodology 
proposed by W\&G to assess the uncertainty in the estimated Bayes partition and present the output of this experiment in Figure~\ref{fig:simulations_2}.
In this case, while the optimal Bayes partition seems very reasonable, having found three contiguous group, the bounds of the credible ball 
appear quite diverse and ``distant'' from the actual optimal solution (particularly the horizontal one). We feel that these bounds do not necessarily convey much 
information regarding which partitions are inside the ball and which are not. %In fact, the primary message seems to be that the uncertainty regarding the clustering is very high. 
Of course, this is a situation where the model is mis-specified, as is the usual case in practice, and this may partially explain the results in Figure~\ref{fig:simulations_2}.

Alternatively, one may instead consider an approach based on the idea of high posterior density regions, and simply list all of the partitions that have posterior 
probability above a certain threshold. This method would include all of the relevant partitions regardless of their distance from the Bayes partition (in the sense of the distance 
induced by the Hasse diagram), providing a good representation of what the possible optimal alternatives look like. From a computational perspective, both methods are 
straightforward to implement once the posterior values and the distances to the Bayes solution are available for all of the partitions sampled. 

\section{Wider application of mixture models}
W\&G propose applications of their methodology to Gaussian mixture models. 
We would like to conclude our discussion by remarking that the method they proposed may be applied in more general mixture modelling contexts, thereby widening 
their applicability. 
For instance, recent research has focused much on mixture models for network data \parencite{daudin2008mixture}.
Computationally efficient Markov chain Monte Carlo sampling strategies for network clustering models have been proposed by \textcite{mcdaid2013improved} and 
\textcite{wyse2012block}. In R\&F, we propose several applications of the decision theoretic framework to Gaussian mixture models, but also to stochastic block models 
for networks, and to latent block models for bipartite networks. Furthermore, mixed-membership models \parencite{airoldi2008mixed} extend the basic clustering structures 
to partial memberships, where nodes of the network may distribute their affiliation among the groups. Extending the decision theoretic framework proposed by 
W\&G to these contexts would be a great next step forward. 

\printbibliography

\end{document}